\documentclass[prl,showpacs,byrevtex,amsmath,twocolumn,amssymb,nofootinbib]{revtex4}
\usepackage{amsmath,graphicx}
\usepackage{epsfig}

\begin{document}

\title{Polymer translocation through a nanopore - a showcase of anomalous
diffusion.}

\author{J. L. A. Dubbeldam$^{1,2}$, A. Milchev$^{1,3}$, V.G. Rostiashvili$^1$ and
T.A. Vilgis$^1$}
\affiliation{$^1$ Max Planck Institute for Polymer Research
  10 Ackermannweg 55128 Mainz, Germany.\\
$^2$ Delft University of Technology 2628CD Delft, The Netherlands\\
$^3$ Institute for Physical Chemistry Bulgarian Academy of Science, 1113 Sofia,
Bulgaria\\
}

\begin{abstract}
The translocation dynamics of a polymer chain through a nanopore in the absence
of an external driving force is analyzed by means of scaling arguments,
fractional calculus, and computer simulations. The problem at hand is mapped
on a one dimensional {\em anomalous} diffusion process in terms of reaction
coordinate $s$ (i.e. the translocated number of segments at time $t$) and shown
to be governed by an universal exponent $\alpha = 2/(2\nu+2-\gamma_1)$ where $\nu$ 
is the Flory exponent, and $\gamma_1$ is the surface exponent. Remarkably, it turns out 
that the value of $\alpha$ is nearly the same in two- and three-dimensions. 
The process is described by a {\em fractional} diffusion 
equation which is solved exactly in the interval $0 <s < N$ with appropriate 
boundary and initial conditions. The solution gives
the probability distribution of translocation times as well as the
variation with time of the statistical moments: $\langle s(t)\rangle$, and
$\langle s^2(t)\rangle  - \langle s(t)\rangle^2$ which provide full description
of the diffusion process. The comparison of the analytic results with data
derived from extensive Monte Carlo (MC) simulations reveals very good agreement
and
proves that the diffusion dynamics of unbiased translocation through a nanopore
is anomalous in its nature. 
\end{abstract}


\maketitle

The dynamics of polymer translocation through a pore has recently received  a
lot of attention and appears highly relevant in both chemical and biological
processes\cite{Meller}. The theoretical cosideration is usually based on the
assumption \cite{Sung1,Sung2,Muthu} that the problem can be mapped onto
a one dimensional diffusion process. The so called {\it translocation
coordinate} (i.e., reaction coordinate $s$) is considered as the only relevant
dynamic variable. The whole polymer chain of length $N$ is assumed as being in
equilibrium with a corresponding free energy ${\cal F} (s)$ of an entropic
nature. The $1 d$-dynamics of the translocation coordinate follows then the
conventional Brownian motion and the one dimensional Smoluchowsky equation
\cite{Risken} can be used with the free energy ${\cal F} (s)$ playing the role
of an external potential. In the absence of external driving force (unbiased
translocation), the corresponding average first - passage time follows the law
$\tau (N) \propto a^2 N^2/D$, where $a$ is the length of a polymer Kuhn segment
and $D$ stands for the proper diffusion coefficient. The question about
the choice of the proper diffusion coefficient, $D$, and the nature of the
diffusion process, is controversial. Some authors \cite{Sung1,Sung2} adopt
$D\propto N^{-1}$, as for Rouse diffusion, which yields $\tau \propto
N^3$ as for polymer reptation\cite{Doi} albeit the short pore constraint is less
severe than that of a tube of length $N$. In ref. \cite{Muthu} it is assumed
that $D$ is not the diffusion coefficient of the whole chain but rather that of
the monomer just passing through the pore. The unbiased translocation time is
then predicted to go as $\tau \propto N^2$. The latter assumption has
been questioned \cite{Kantor1,Kantor2} too. Indeed, on the one hand, the mean
translocation time scales\cite{Muthu} as $\tau \sim N^2$ but on the other hand
the characteristic Rouse time (i.e. the time it takes for a {\em free} polymer
to diffuse a distance of the order of its gyration radius) scales as $
\tau_{\rm Rouse}\propto N^{2\nu+1}$ where the Flory exponent $\nu=0.588$ at
$d=3$, and $\nu=0.75$ at $d=2$ \cite{Doi}. Thus $\tau_{\rm Rouse} \gg \tau$
against common sense given that the unimpeded motion should be in any way
faster than that of a constrained chain! Moreover, the equilibration of the
chain is questionable when the expression for ${\cal F} (s)$ is to be used. The
characteristic equilibration time scales again as $\tau_{\rm eq} \propto
N^{2\nu+1}$ and is thus always larger than the translocation time, i.e.
$\tau_{\rm eq} \gg \tau$. Again the internal consistency of the whole approach
is in doubt. It was found by MC-simulation \cite{Kantor1,Kantor2} that
$\tau \propto N^{2.5}$ for translocations in $d=2$. This indicates that the
translocation time scales in the same manner as the Rouse time albeit with a
larger prefactor that depends on the size of the nanopore. The authors argued
that this finding bears witness of the failure of the Brownian nature of the
translocation dynamics and suggested instead that {\it anomalous
diffusion dynamics} \cite{Metzler} should be more adequate. The $\tau \propto
N^{2.5}$ scaling law has been corroborated by a further MC - study
\cite{Luo} as well as by a MC-simulations on a $3 d$ - lattice \cite{Barkema},
and it was shown that $\tau \propto N^{2.46\pm 0.03}$. The time variation of the
second statistical moment, $\left\langle s^2 \right\rangle - \left\langle
s\right\rangle ^2 \propto t^{\alpha}$, clearly indicates an anomalous nature
\cite{Barkema} since the measured exponent $\alpha = 0.81\pm 0.01$ while 
$\tau \propto N^{2/\alpha}$. Still missing is a proper theoretical
analysis which could explain the physical origin of the anomalous
dynamics, and make it possible to solve the appropriate {\em fractional
diffusion equation} (or, in case of a biased translocation, the fractional Fokker -
Planck - Smoluchowski equation) \cite{Metzler,Metzler1} governing this
dynamics. 

In this Letter we suggest a unique physical picture which justifies the mapping of the
$3 d$ problem on a $1 d$ reaction cooordinate, $s$, and we show
that the latter obeys anomalous diffusion dynamics, described by a fractional
diffusion equation. We solve this equation exactly, subject to the proper
boundary conditions, and find a perfect agreement with our scaling prediction.
Eventually, we demonstrate that the results of our $3 d$ off-lattice
MC-simulations are in accord with our analytical findings.

\paragraph*{Mapping onto $1 d$ dynamics} - As indicated above, the assumption
that the whole polymer chain is in equilibrium and the diffusion is
governed by conventional Brownian dynamics leads to contradictions. Instead,
we assume now that only a {\em part} of the whole chain may equilibrate
between two successive threadings. This part of the chain which adjoins the membrane on the {\it cis}-  or  {\it trans}- sides will be denoted as
{\it fold} and we assume that it is much shorter than the whole chain length $N$
but is still long enough so that one can use the principles of statistical
physics. We also assume that the excluded volume interaction of a fold with the
rest of the chain is relatively weak so that it could be treated as a subsystem
with a well defined free energy.  This latter assumption is based on the
observation that the chain on either of the two sides may be seen as
a polymeric "mushroom" whereby the monomer density close to the
membrane (or wall) is much smaller than the density inside a single coil (see
Fig.4 and eq.(II.4) in ref.\cite{Gennes1}). Thus one can claim that there is a
depletion area near the membrane \cite{Gennes2}

\begin{figure}
\includegraphics[scale=0.3]{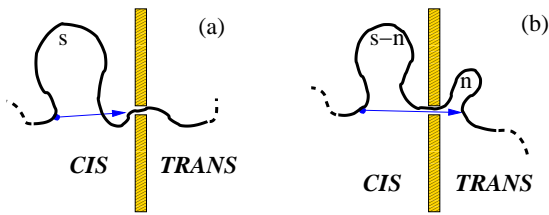}
\caption{Schematic representation of a chain fold of length $s$ moving through a
nanopore. The transition rate is slowed down by an entropic barrier: a) 
initially the fold is on the {\it cis}-side of the wall, b) the fold entropy
decreases during threading because of the fold fragmentation.}
\label{Fold}
\end{figure}

Figure \ref{Fold} illustrates how a fold squeezes from the {\it cis}- to the
{\it trans} - side  through a short nanopore (of length $\approx a$) which is
slightly wider than the chain itself. 
%\onecolumngrid
%\twocolumngrid
It is self - evident that, in the absence of the external force, with the equal probability folds from the {\it trans}- side could go to the {\it cis} - side.
If the {\it trans}-part of the fold in Fig.\ref{Fold} has length $n$ then the corresponding
free energy function $F^{t} (n)/T =  -n \ln \kappa - (\gamma_1  - 1) \ln n $,
where $\kappa$ is the connective constant and $\gamma_1$ is the surface
entropic exponent \cite{Vander}. For the {\it cis}-part of the fold one has 
$F^{c} (n)/T = -(s - n) \ln \kappa - (\gamma_1  - 1) \ln (s -
n)$ so that the total free energy is ${\cal F} (n)/T =  - s\ln \kappa -
(\gamma_1  - 1) \ln [n(s-n)]$. One can, therefore, ascribe to the fold {\it 
cis-trans} - transition a pretty broad barrier given by ${\cal F} (n)$. The
corresponding activation energy of the fold can be calculated as $\Delta E (s) =
{\cal F} (s/2) - {\cal F} (1) = (1 - \gamma_1) T \ln s$. 

How can we estimate the characteristic time for the fold {\it cis - trans} -
translocation? In the absence of a separating membrane this would be the pure
Rouse time $t_{\rm R} \propto s^{2\nu + 1}$  \cite{Doi}. The membrane with a nanopore imposes
an additional entropic activation barrier $\Delta E (s)$ which slows down the
transition rate.
The characteristic time, therefore, scales as $t(s) = t_{\rm R} (s) \exp
[\Delta E (s)] \propto s^{2\nu + 2 - \gamma_1}$. This makes it possible to
estimate the mean-squared displacement of the $s$ - coordinate:
\begin{equation}
\left\langle s^2 \right\rangle \propto t^{2/(2\nu + 2 - \gamma_1)} .
\end{equation}
Hence the mapping onto the $s$ - coordinate leads to an {\em anomalous
diffusion} law, $\left\langle s^2\right\rangle \propto t^{\alpha}$, where
$\alpha = 2/(2\nu + 2 - \gamma_1)$. Taking into account the most accurate 
values of the  exponents for $d = 3$ , $\nu = 0.588$ and $\gamma_1 = 0.680$
\cite{Diehla}, we obtain  $\alpha = 0.801$. In turn, the average translocation
time $\tau \propto N^{2/\alpha} \propto N^{2.496}$. Remarkably, in $2 d$
where $\nu_{2d}=0.75$ and $\gamma_1\approx 0.945$\cite{Whittington}, one finds
$\alpha\approx 0.783$ (i.e., $\alpha$ is almost unchanged)! This explains why 
the measured exponents in both $2d$ \cite{Kantor1} and in $3d$ \cite{Barkema} 
are so close. The derivation of $\alpha$  is our central scaling prediction. It
also agrees well (see below) with our own MC - data on the translocation exponent.

\paragraph*{Fractional diffusion equation} - We now turn to the fractional
diffusion equation (FDE)  which furnishes a natural framework for the study of
anomalous diffusion \cite{Metzler,Metzler1}. Here we make use of this method in
a systematic way. Our FDE reads
\begin{equation}
\frac{\partial}{\partial t} \:  W (s, t) = {_0}D_{t}^{1 - \alpha} K_{\alpha}
\frac{\partial^2}{\partial s^2} \: W (s, t) \quad,
\label{FDE}
\end{equation}
where $W (s, t)$ is the probability distribution function (PDF) for having a segment
$s$ at time $t$ in the pore, and the fractional Riemann - Liouville operator
${_0}D_{t}^{1 - \alpha} W (s, t) = (1/\Gamma (\alpha)) (\partial/\partial t)
\int_0^t d t' W (s, t')/(t - t')^{1-\alpha}$. In Eq. (\ref{FDE}) $\Gamma (\alpha) $ is the
Gamma-function, and $K_{\alpha}$ is the so called generalized diffusion
constant. This constant could be defined as $K_{\alpha}=\Gamma(1+\alpha)l^2/(2\tau_{\rm w}^{\alpha})$ in terms of the fold length $l$ and the waiting time scale $\tau_{\rm w}$ (see Chapter 3.4 in \cite{Metzler}).  
It should be mentioned that the constant $K_{\alpha}$ is the only adjustable parameter of our theory which will be fixed below through the comparison with our MC - data. 

Recently the method of generalized Langevin equation (GLE) has been used to
describe anomalous conformational dynamics within single molecule proteins
\cite{Debnath}. In contrast to the FDE - approach which deals with the total
distribution function at particular boundary conditions (see below), the GLE
- method treats only the first two moments (or time-correlation functions,
memory kernel, etc.).  To the best of our knowledge, at the present time it is
not clear how one can derive in a closed form a
non-Markovian Fokker-Planck equation for the distribution function
\cite{Coffey} starting from the GLE.
On the other hand,  the translocation time distribution function (see below) is
an entity of great importance because it could be directly measured in
experiment \cite{Meller}. Therefore, we prefer to use FDE - approach for the
translocation problem.

Consider the boundary value
problem for FDE in the interval $0\leq s \leq N$. This problem has been discussed before in the context of even more general fractional Fokker-Planck equation (FFPE) \cite{Metzler2}. The boundary conditions
correspond to the {\em reflecting-adsorbing} case, i.e., $(\partial/\partial s)
W (s, t)|_{s=0} = 0$ and $W (s = N, t) = 0$. The initial distribution is
concentrated in $s_0$, i.e., $W (s, t=0) = \delta (s - s_0)$. The full solution
can be represented as a sum over all eigenfunctions $\varphi_n (s)$, i.e., $W
(s, t) = \sum_{n=0}^{\infty} T_n (t) \varphi_n (s)$ where $\varphi_n (s)$ obey
the equations $K_{\alpha} (d^2/ds^2) \varphi_n (s) + \lambda_{n,\alpha}\varphi_n
(s) = 0 $, and the eigenvalues $\lambda_{n,\alpha}$ can be readily found from
the foregoing boundary conditions, as a result $\lambda_{n,\alpha}=(2n+1)^2\pi^2K_{\alpha}/(4N^2)$.
The temporal  part $T_n (t)$ obeys the
equation $(d/d t) T_n (t) = - \lambda_{n,\alpha} \: {_0}D_{t}^{1 - \alpha} T_n
(t)$. The solution of this equation is given by $T_n (t) = T_n (t=0) E_{\alpha} (-
\lambda_{n,\alpha} \: t^{\alpha})$ \cite{Metzler} where the Mittag - Leffler
function $E_{\alpha}(x)$ is defined by the series expansion $E_{\alpha} (x) =
\sum_{n=0}^{\infty} \: x^k/\Gamma (1+\alpha k)$. At $\alpha = 1$ it turns
back into a standard exponential function (normal diffusion). Thus we arrive
at the complete solution of eq. ( \ref{FDE}):
\begin{eqnarray}
W (s, t) &=& \frac{2}{N} \:  \sum_{n = 0}^{\infty} \:\cos\left[ \frac{(2 n + 1)
\pi s_0}{2 N}\right] \: \cos\left[ \frac{(2 n + 1) \pi s}{2 N}\right]
\nonumber\\
&\times& E_{\alpha}\left[ - \frac{(2 n + 1)^2 \pi^2}{4  N^2} \: K_{\alpha} \:
t^{\alpha}\right] \quad.
\label{Solution_box}
\end{eqnarray}
\paragraph*{First - passage time distribution} - The distribution of
translocation times (which could, in principle, be measured in experiment) is
nothing but the {\it first - passage time  distribution}  (FPTD)  $Q (s_0, t)$,
where $s_0$  stands for the initial value of the $s$ - coordinate. Knowing the
probability distribution function $W (s, t)$, we can calculate FPTD  $Q (s_0,
t)$. The relation  between both functions is given as $Q (s_0, t) =  - (d/d t)
\int_{0}^t \: W (s, t) ds$ \cite{Risken}. This yields the FPTD as follows 
\begin{eqnarray}
Q(s_0, t) &=& \frac{\pi K_{\alpha} t^{\alpha - 1}}{N^2} \: \sum_{n = 0}^{\infty}
\:(-1)^n (2 n + 1) \: \cos\left[ \frac{(2 n + 1) \pi s_0}{2 N}\right]
\nonumber\\
 &\times& E_{\alpha, \alpha}\left[ - \frac{(2 n + 1)^2 \pi^2}{4 N^2} \:
K_{\alpha} \:  t^{\alpha}\right] \quad,
\label{Final_time}
\end{eqnarray}
where the generalized Mittag - Leffler function $ E_{\alpha, \alpha} (x) = 
\sum_{k = 0}^{\infty} \: x^k/\Gamma (\alpha + k \alpha)$. The long time limits 
of Mittag - Leffler functions in eqs. (\ref{Solution_box}) and
(\ref{Final_time})  follow an inverse power law behavior, $E_{\alpha} (-
\lambda_{n, \alpha} t^{\alpha}) \propto 1/\Gamma (1- \alpha) \lambda_{n, \alpha}
t^{\alpha}$ and $E_{\alpha, \alpha} (- \lambda_{n, \alpha} t^{\alpha}) \propto
\alpha/\Gamma (1- \alpha) \lambda_{n, \alpha}^2  t^{2 \alpha}$. By making use of
this in eq. (\ref{Final_time}), the long time tail of the FPTD then reads $Q (t)
\propto \alpha N^2/2\Gamma (1-\alpha)K_{\alpha} t^{1+\alpha}$. This behavior
is checked below in our MC - investigation. It can be seen that the mean
first passage time, defined simply as $\tau = \int_{0}^{\infty} t Q (t) d t$, 
does not exist \cite{Grosberg,Katja}. On the other hand, in a laboratory experiment
there always exists some upper time limit $t^{\ast}$. Taking this into account,
one can show that an "experimental" first passage time scales as
$\tau \sim N^{2/\alpha}$ \cite{Grosberg} which we observe in our MC-simulation.

\paragraph*{Statistical moments $\langle s\rangle $ and
$\langle s^2 \rangle $ vs. time} -  The subdiffusive behavior of the
second moment  $\langle s^2 \rangle - \langle s \rangle^2
\propto t^{\alpha}$ is a hallmark of anomalous diffusion. Starting from eq.
(\ref{Solution_box}) we can immediately calculate them. The calculation of the
first moment $\left\langle s \right\rangle = \int_{0}^{N} s W (s, t) ds /
\int_{0}^{N} W (s, t) ds$ yields
\begin{eqnarray}
\frac{\left\langle s \right\rangle  (t)}{N} = 1-\frac{2 \sum_{n = 0}^{\infty} \:
\frac{1}{(2n+1)^2} \: E_{\alpha}\left[ - \frac{(2 n + 1)^2 \pi^2}{4  N^2} \:
K_{\alpha} \: t^{\alpha}\right] }{\pi \sum_{n = 0}^{\infty} \:
\frac{(-1)^n}{(2n+1)} \: E_{\alpha}\left[ - \frac{(2 n + 1)^2 \pi^2}{4  N^2} \:
K_{\alpha} \: t^{\alpha}\right]}  
\label{First_moment_final}
\end{eqnarray}
Since $E_{\alpha} (t=0) = 1$, the initial value $\langle s \rangle (t=0) = 0$
(we put $s_0 = 0$) as it should be. In the opposite limit, $t \rightarrow \infty$, we can use the asymptotic behavior $E_{\alpha}[- \lambda_{n,\alpha} t^{\alpha}] \simeq 1/\Gamma(1-\alpha)\lambda_{n,\alpha}t^{\alpha}$ as well as the sums values $\sum_{n=0}^{\infty} 1/(2n+1)^4= \pi^4/96$ and $\sum_{n=0}^{\infty} (-1)^n/(2n+1)^3 = \pi^3/32$ in the nominator and denominator respectively. After that $\langle s \rangle 
(t \rightarrow \infty) = N/3$, i.e. the function goes to a {\em plateau}. 

The
result for the second moment, $\left\langle s^2 \right\rangle = \int_{0}^{N} s^2
W (s, t) ds / \int_{0}^{N} W(s, t) ds$, can be cast in the following form
\begin{eqnarray}
\frac{\left\langle s^2 \right\rangle  (t)}{N^2}=1-\frac{8 \sum_{n = 0}^{\infty}
\: \frac{(-1)^n}{(2n+1)^3} \: E_{\alpha}\left[ - \frac{(2 n + 1)^2 \pi^2}{4 
N^2} \: K_{\alpha} \: t^{\alpha}\right] }{\pi^2 \sum_{n = 0}^{\infty} \:
\frac{(-1)^n}{(2n+1)} \: E_{\alpha}\left[ - \frac{(2 n + 1)^2 \pi^2}{4  N^2} \:
K_{\alpha} \: t^{\alpha}\right]}
\label{Second_moment}
\end{eqnarray}
Again it can be readily shown that $\langle s^2 \rangle (0) -\langle s \rangle^2
 (0) = 0$. In the long time limit in the same way as before and taking into account that $\sum_{n=0}^{\infty} (-1)^n /(2n+1)^5 = 5\pi^5/1536$ we will find $\left\langle s^2 \right\rangle
(t\rightarrow \infty) - \left\langle s \right\rangle^2  (t\rightarrow \infty) =
N^2/9$ \cite{Note}.

\paragraph*{Monte Carlo data vs. theory} - We have carried out extensive MC
- simulations in order to check the main predictions of the foregoing analytical
theory. We use a dynamic bead-spring model which has been described before
\cite{Milchev}, therefore we only mention the salient features here. Each
chain contains $N$ effective monomers (beads), connected by anharmonic
FENE (finitely extensible nonlinear elastic) springs, and the nonbonded segments
 interact by a Morse potential. An elementary MC move is performed by picking an
effective monomer at random and trying to displace it from its position
to a new one chosen at random. These trial moves are
accepted as new configurations if they pass the standard Metropolis acceptance
test. It is well established that such a MC algorithm, based on local moves,
realizes Rouse model dynamics for the polymer chain. In the course of the
simulation we perform successive run for chain lengths $N=16, 32,64, 128, 256$
whereby a run starts with a configuration with only few segmens on the
trans-side. Each run is stopped, once the entire chain moves to the
trans-side. Complete retracting of the chain back to the cis-side is
prohibited. During each run we record the translocation time $\tau$, and the
translocation coordinate $s(t)$. Then we average all data over typically $10^4$ runs. In
\begin{figure}
\includegraphics[scale=0.35,angle=270]{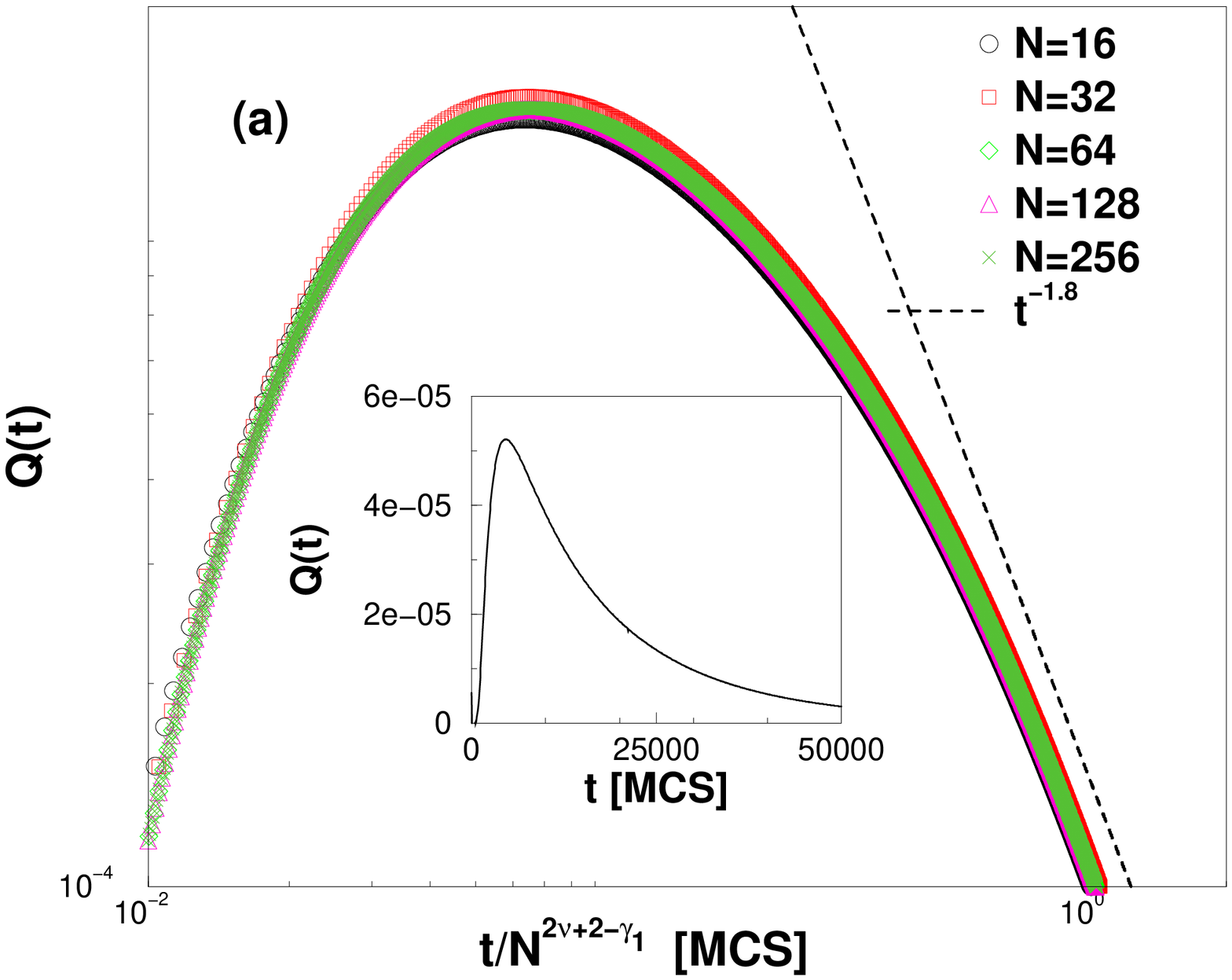}
\includegraphics[scale=0.35,angle=270]{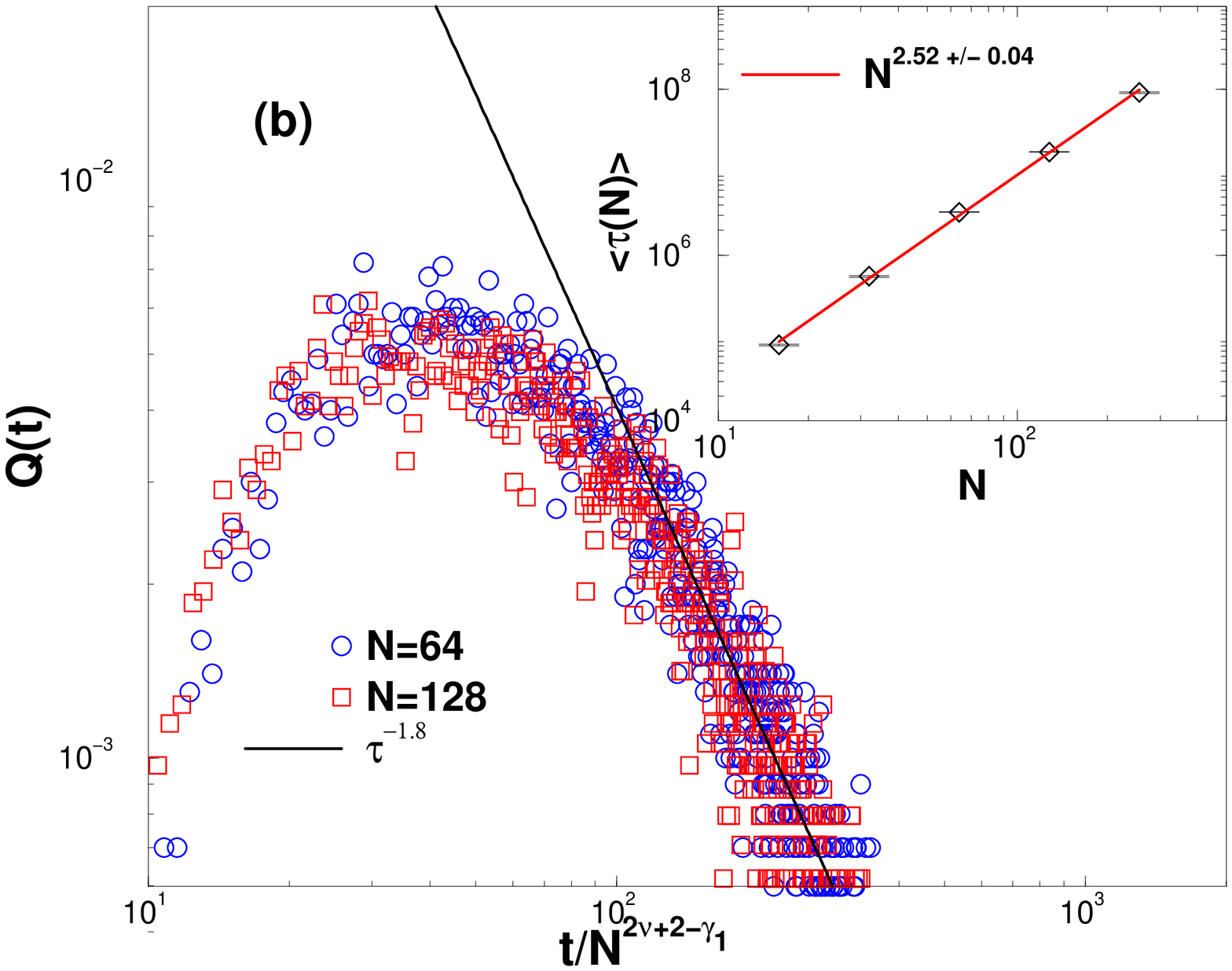}
\caption{The translocation time distribution function $Q (t)$: a) Scaling
plot of the theoretical predictions calculated from Eq.
(\ref{Final_time}) for different chain lengths $N$. Dashed line denotes the
long time asymptotic tail with slope $-1.8$. The inset shows $Q(t)$ for $N=256$
in normal coordinates. b) The FPDT $Q(t)$ from the MC-simulation for $N=64, 
128$. The inset shows the expected $\langle\tau\rangle$ vs. chain length $N$
dependence and the straight line is a best fit with slope $\approx 2.52 \pm
0.04$.}
\label{PDF}
\end{figure}
principle, the pore may apply a drag force on the threading chain due to a
chemical potential gradient, however, in the present work we consider only
unbiased diffusion. In Figure \ref{PDF} (a) we show a master plot of the
translocation time distribution $Q(t)$ derived from Eq. (\ref{Final_time})
for different chain lengths $N=16, 32, 64, 128, 256$. For the calculation 
of data we have used {\em Mathematica} with a special package for
computation of Mittag-Leffler functions\cite{Loutchko}. Evidently, all curves
collapse on a single one when time is scaled as $t\propto N^{2/\alpha}$ with the
predicted $\alpha = 0.8$. The long-time tail for this value of $\alpha$ should
exhibit a slope of $-1.8$. The inset in Figure \ref{PDF}(a) reveals the long
tail of $Q(t)$ for large times. A comparison with Figure \ref{PDF}(b)
demonstrates good agreement with the simulation data despite some scatter in the
FPDF even after averaging over $10000$ runs. As shown in the inset, the mean
translocation time scales as $\langle\tau\rangle \propto N^{2.5}$ in good
agreement with the predicted $\alpha = 0.8$. 
\begin{figure}
\includegraphics[scale=0.4,angle=270]{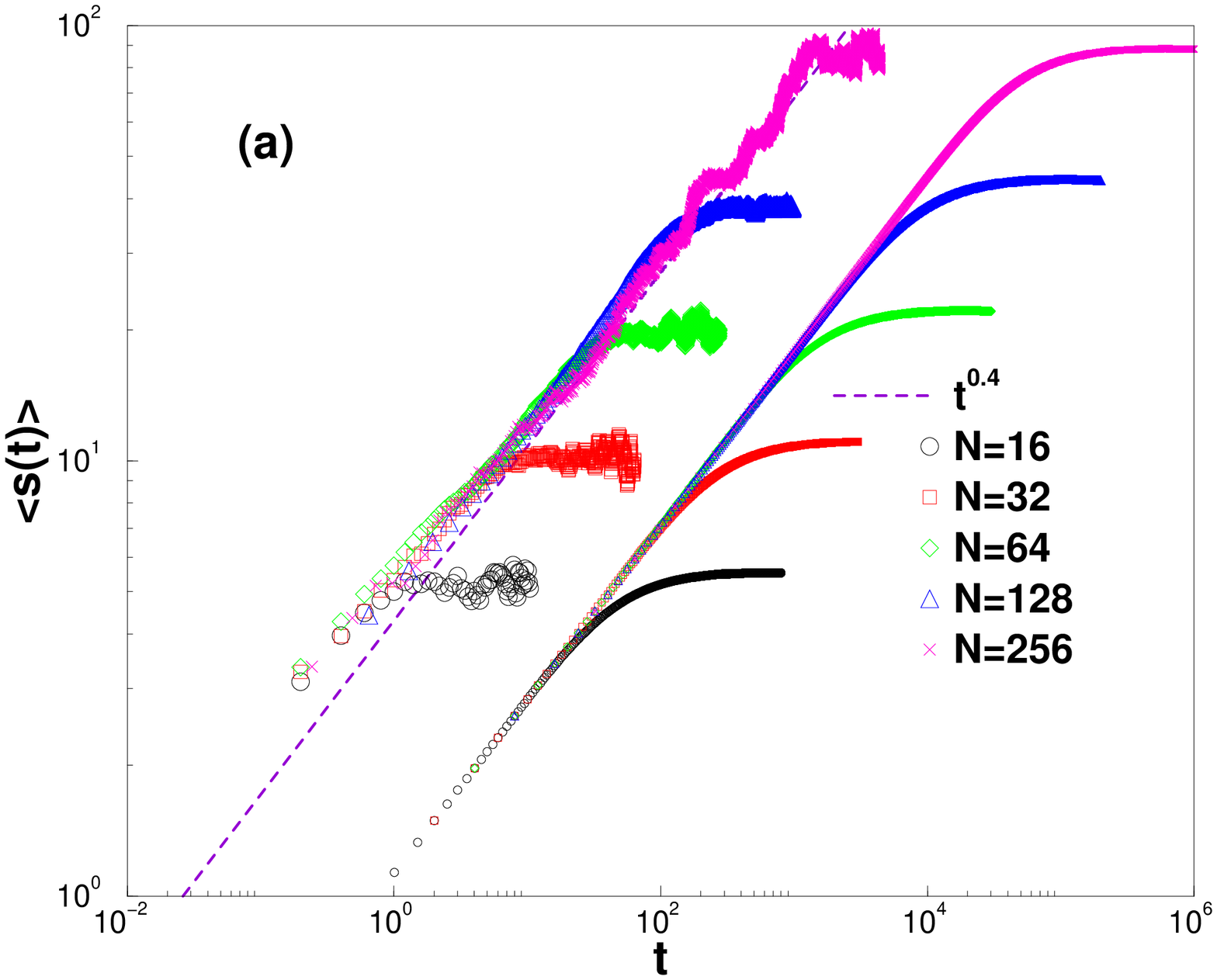}
\includegraphics[scale=0.4,angle=270]{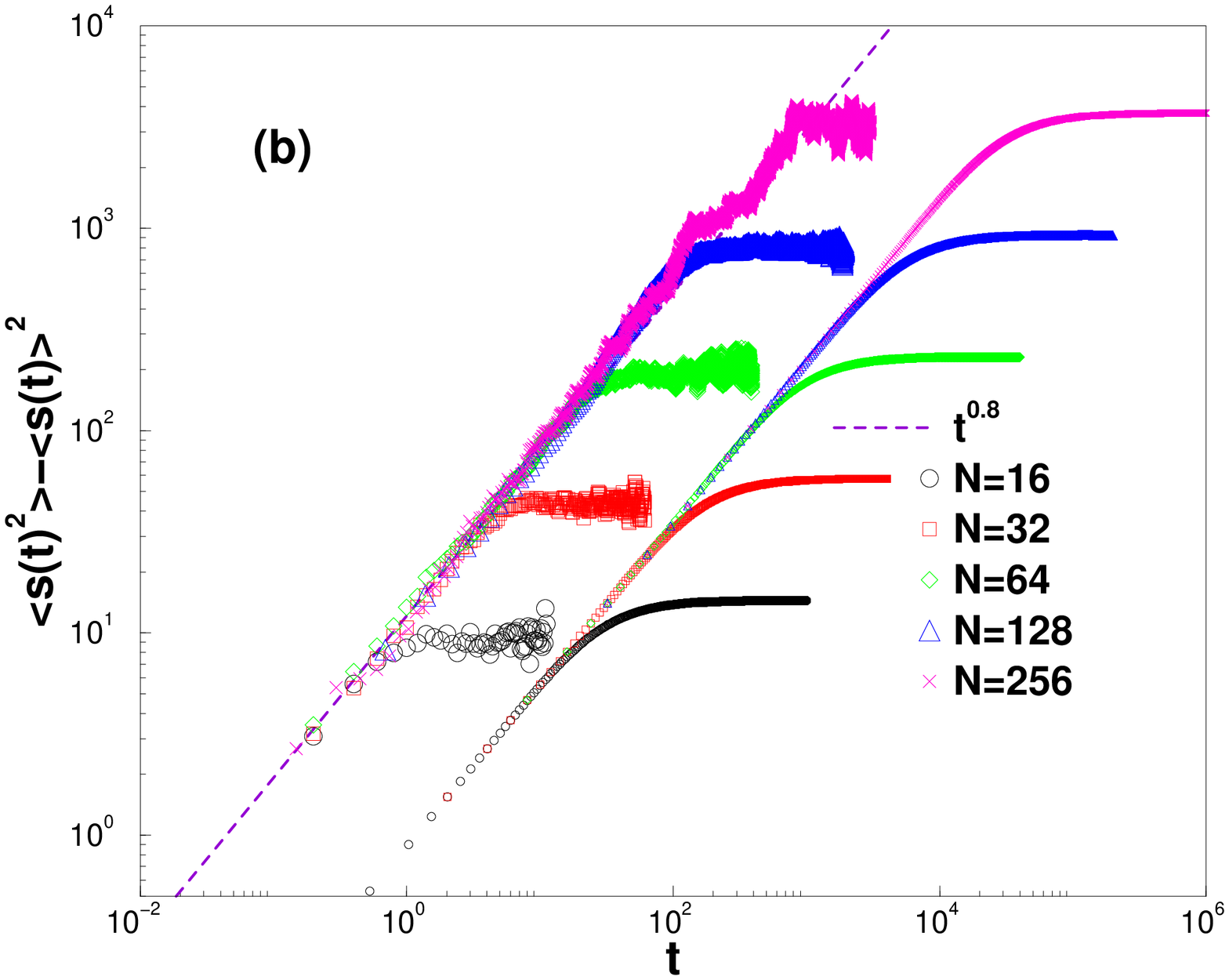}
\caption{Variation of the first and second moments of the PDF $W(s, t)$  with time for
chain lengths $N=16, 32, 64, 128, 256$. a) Log-log plot of the first moment $\langle
s\rangle$ vs. time $t$ from a MC simulation (big symbols) and from Eq.
(\ref{First_moment_final}) (small symbols). The dashed line denotes
$t^{\alpha/2}$ with a slope of $0.4$. b) The same as in a) but for 
 $\langle s^2\rangle - \langle s\rangle ^2$. Analytical data is
obtained from Eq. (\ref{Second_moment}). The dashed line has a slope of $0.8$.}
\label{Moments}\textit{}
\end{figure}
An inspection of Figure \ref{Moments} where the time variation of the
PDF $W(s,t)$  moments are compared demonstrates again that data from the numeric
experiment and the analytic theory agree well within the limits of statistical
accuracy (which is worse for $N=256$). Not surprisingly, the time scale of
the MC results does not coincide with that of Eqs.
(\ref{First_moment_final}), (\ref{Second_moment}) since in the latter we have
set $K_{\alpha}$, which fixes the time scale, equal to unity. Closer examination of Figure \ref{Moments} shows that the resetting of the generalized diffusion coefficient as $K_{\alpha} \simeq (80)^{0.8} \simeq 33.3$ enables to superimpose the results of theoretical calculation and MC - data.

In summary we have shown unambiguously that the translocation dynamics of a
polymer chain threading through a nanopore is anomalous in its nature. We have
succeeded to calculate the anomalous exponent $\alpha = 2/(2\nu+2-\gamma_1)$
from simple scaling arguments and embedded it in the fractional diffusion
formalism. We derived exact analytic expressions for the translocation time
probability distribution as well as for the moments of the translocation
coordinates which are shown to agree well with our MC simulation data. The
present treatment can be readily generalized to account
for a drag force on the chain and results for this case will be reported
in a separate publication.

\paragraph*{Acknowledgments} - 
The authors are indebted to Burkhard D\"unweg for stimulating discussion during
this study. The permission to use the special package for computation of
Mittag-Leffler functions with {\em Mathematica}, provided by Y. Luchko, is
gratefully acknowledged. The authors are indebted to $SFB-DFG 625$ for
financial support during this investigation.


\begin{thebibliography}{99} 
\bibitem{Meller} A. Meller, J. Phys. Condensed Matter {\bf 15}, R581(2003).
\bibitem{Sung1} W. Sung, P.J. Park, Phys. Rev. Lett. {\bf 77}, 783 (1996).
\bibitem{Sung2} P.J. Park, W. Sung, J. Chem. Phys. {\bf 108}, 3013 (1998).
%\bibitem{Sung3} P.J. Park, W. Sung, Phys. Rev.E {\bf 57}, 730 (1998).
\bibitem{Muthu} M. Muthukumar, J. Chem. Phys. {\bf 111}, 10371 (1999).
\bibitem{Risken} H. Risken, {\it The Fokker - Planck Equation}, Springer -
Verlag, Berlin , 1989.
\bibitem{Kantor1} J. Chuang, Y. Kantor, M. Kardar, Phys. Rev. E {\bf 65}, 011802
(2001).
\bibitem{Kantor2}Y. Kantor, M. Kardar, Phys. Rev. E {\bf 69}, 021806 (2004).
\bibitem{Doi} M. Doi, S.F. Edwards, {\it The Theory of Polymer Dynamics}
(Clarendon Press, Oxford, 1986).
\bibitem{Metzler} R. Metzler, J. Klafter, Physics Rep. {\bf 339}, 1 (2000)
\bibitem{Luo}K. Luo, T.Ala-Nissila, S-C. Ying, J. Chem. Phys. {\bf 124}, 0334714
(2006).
\bibitem{Barkema} D. Panja, G. Barkema, R.C. Ball, arXiv:cond-mat/0610671.
\bibitem{Metzler1} R. Metzler, J. Klafter,  Biophys. Journal, {\bf 85}, 2776
(2003).
\bibitem{Gennes1} P.G. de Gennes, Macromolecules {\bf 13}, 1069 (1980).
\bibitem{Gennes2} P.G. de Gennes,  Adv. Colloid and Interface Sci. {\bf 27}, 189 (1987).
\bibitem{Vander} C. Vanderzande, {\it Lattice Models of Polymers} (Cambridge
University Press, Cambridge, 1998).
\bibitem{Diehla} H.W. Diehla, M. Shpot, Nucl. Phys. B {\bf 528}, 595 (1998); R.
Hegger, P. Grassberger, J. Phys. A {\bf 27}, 4069 (1994).
\bibitem{Whittington} M.N. Barber, A.J. Guttmann, K.M. Middlemiss, G.M. Torrie, S.G. Whittington, J. Phys. A {\bf 11}, 1833 (1978).
\bibitem{Debnath} P. Debnath, W. Min, X.S. Xie, B.J. Cherayil, J. Chem. Phys. {\bf 123}, 204903 (2005); W. Min, G. Luo, B.J. Cherayil, S.C. Kou, X.S. Xie, Phys. Rev. Lett. {\bf 94}, 198302 (2005); S.C. Kou, X.S. Xie, Phys. Rev. Lett. {\bf 93}, 180603 (2004).
\bibitem{Coffey} W.T. Coffey, Yu.P. Kalmykov, J.T. Waldron, {\it The Langevin Equation}, World Scientific Publishing, London, 2004.
\bibitem{Metzler2} R. Metzler, J. Klafter, Physica A {\bf 278}, 107 (2000).
\bibitem{Grosberg} R.C. Lua, A.Y. Grosberg, Phys. Rev. E {\bf 72}, 061918
(2006).
\bibitem{Katja} S.B. Yuste, K. Lindenberg, Phys. Rev. E {\bf 69}, 033101 (2004).
\bibitem{Note} It comes as no surprize that the prefactors less than unity appear in $\left<s\right> (\infty)$ and $\left<s^2\right>(\infty)$. For example $\left<s^2\right>(\infty)=\int_{0}^{N}s^2 W(s, \infty)ds/\int_{0}^{N}W(s, \infty)ds = \xi^2 N^2$, where $0<\xi<1$ and we have used the well known {\it mean value theorem} for integration (see any integral calculus textbook). This also does not contradict to the condition which fixes the "experimental" average first passage time $\tau$: $\left<s^2\right>(t=\tau) \sim N^2$ i.e. $\tau \sim N^{2/\alpha}$ \cite{Grosberg} . From the physical perspective this means that the particular observation where $s^2(t=\tau) = N^2$ is a relatively rear event in the total sampling.
\bibitem{Milchev} A. Milchev, K. Binder and A. Bhattacharya, J. Chem. Phys.
{\bf 121}, 6042(2004).
\bibitem{Loutchko} R. Gorenflo, J. Loutchko, and Y. Luchko, Fractional
Calculus and Applied Analysis, {\bf 5}, 491(2002).
\end{thebibliography}
\end{document}